\documentclass[trackchanges,twocolumn]{aastex701}
\usepackage{amsmath}
\usepackage{enumerate}

\begin{document}

\title{Anisotropic Ejecta from Binary Neutron Star Mergers: Self-Consistent Main Thermal and Late-Time Radio Emission of NS-Powered Kilonovae}

\author{Jia-Xiang Chen}
\affiliation{College of Physical Science and Technology, Hebei
University, Baoding 071002, China}
\affiliation{Hebei Key
Laboratory of High-precision Computation and Application of Quantum
Field Theory, Baoding, 071002, China}
\affiliation{Research
Center for Computational Physics of Hebei Province, Baoding, 071002,
China}
\email[]{lishaoze@mails.ccnu.edu.cn}

\author{Shao-Ze Li}
\affiliation{College of Physical Science and Technology, Hebei
University, Baoding 071002, China}
\affiliation{Hebei Key
Laboratory of High-precision Computation and Application of Quantum
Field Theory, Baoding, 071002, China}
\affiliation{Research
Center for Computational Physics of Hebei Province, Baoding, 071002,
China}
\email[show]{lishaoze@mails.ccnu.edu.cn}

\correspondingauthor{Shao-Ze Li}

\begin{abstract}

The interaction between the fast-moving ejecta and the interstellar medium can produce
long-lasting radio signals after binary neutron star mergers.
Searching for such radio signals is a way to test the central engine of
kilonovae and short gamma-ray bursts. With a magnetar as the central engine, the
spin-down energy powers the main thermal and 
late-time radio emissions of the kilonova. However, both the thermal and radio
emissions are strongly affected by the ejecta distribution, e.g., the 
two-component ``blue" and ``red" emissions of AT\,2017gfo corresponding to the GW\,170817 event. In this study, we
investigate the distribution of the merger ejecta, analyzing several
possible anisotropic distributions and demonstrating their impacts on
the emission properties, particularly the late-time radio light
curves. Under a bipolar and equatorial ejecta configuration,
corresponding to the wind and dynamical components of the merger ejecta,
the late-time radio light curves reveal distinct two-peak features, 
which are consistent with the main thermal light curves.
The anisotropic distribution of the ejecta intrinsically connects the
main thermal and late-time radio emissions, forming a
self-consistent evolutionary picture. A combined analysis of the
main thermal and late-time radio emissions provides a way to
constrain the geometry of the merger ejecta and to probe the properties
of the central engine. Furthermore, using the fitting parameters from the main thermal emission of AT\,2017gfo, 
we calculate the corresponding potential late-time radio light curves. The results show that, under typical parameters, 
the non-detection of radio signals in observations is consistent with the theoretical expectation.

\end{abstract}



\section{Introduction} \label{sec:intro}

In 2017, the LIGO and Virgo gravitational-wave detectors announced the first detection of gravitational waves from a binary neutron star (NS) merger \citep[GW\,170817,][]{2017PhRvL.119p1101A}. 
Almost simultaneously, Fermi observed the corresponding short gamma-ray burst, GRB\,170817A \citep{2017ApJ...848L..14G}.
Subsequently, optical and infrared telescopes detected the associated kilonova, AT\,2017gfo \citep{2017Sci...358.1556C}.
This was the first joint observation of gravitational waves, gamma-ray bursts, and kilonovae, confirming the direct connection between binary NS mergers, short gamma-ray bursts, 
and kilonovae \citep{2017ApJ...848L..17C,2017ApJ...848L..18N,2017Natur.551...75S,2017ApJ...848L..27T,2017ApJ...851L..21V}, which heralded a new era in multi-messenger astronomy.

The nature of the remnant of binary NS mergers is one of the most fundamental questions in current astrophysical research. 
It would provide strong constraints on the maximum mass of NSs and the equation of state of dense matter if the merger remnant were
a massive NS rather than a black hole \citep{2019AIPC.2127b0025L, 2020ApJ...893..146A, 2020ApJ...890...99L}. 
A direct method to distinguish the merger product is the detection of post-merger gravitational-wave signals.  If the remnant is a massive NS, it would emit persistent or secular gravitational-wave signals on longer timescales \citep{2019ApJ...875..160A, 2025ApJ...986...14H}. However, the current sensitivity of gravitational-wave detectors is insufficient to detect such gravitational-wave emission.
As an alternative, electromagnetic counterparts to GW events are expected to provide additional observational signatures to determine the properties of the merger remnant. 
The presence of a central NS enables continuous energy injection and powers the GRB afterglow \citep{1998A&A...333L..87D, 1998PhRvL..81.4301D, 2001ApJ...552L..35Z, 2006ApJ...642..354Z}. The observed plateau and extended emission in GRB afterglows indicate the long-lasting activity of the central engine \citep{2013ApJ...763L..22Z, 2015ApJ...805...89L}. The main thermal emission of kilonovae powered by the central NS
is expected to be much brighter than that powered by the radioactive decay \citep{2013ApJ...776L..40Y,2014MNRAS.441.3444M}.  Moreover, the energy injection simultaneously accelerates the merger ejecta, enhancing the interaction between the ejecta and the interstellar medium (ISM) and resulting in a long-lasting radio signal several years after the merger \citep{2013ApJ...771...86G, 2014MNRAS.437.1821M, 2016ApJ...831..141F}. 
However, such a long-timescale radio signal has not been detected in current radio observations \citep{2021MNRAS.500.1708R,2022JApA...43...66G}. 
On the contrary, the radio signals may be overestimated for several plausible reasons. The gravitational-wave radiation probably carries away most of the NS's energy, leaving only a fraction to accelerate the ejecta \citep{2016PhRvD..93d4065G}. The surrounding ISM is incompletely swept up by the merger ejecta \citep{2024ApJ...961..201L}. Furthermore, the emission could be significantly affected by the anisotropic distribution of the anisotropic ejecta, as seen in the two-component kilonova emission of AT\,2017gfo \citep[e.g.,][]{2017Natur.551...75S, 2017ApJ...848L..27T}. 

Therefore, this study focuses on the anisotropic ejecta distribution and investigates the potential influence on the main thermal and late-time radio emissions of kilonovae. The ejecta of binary NS mergers are primarily classified into two distinct components \citep{2020ARNPS..70...95R}: the tidal-force-induced dynamical ejecta, concentrated in the equatorial plane; and the polar wind, launched by shock heating or by accretion of the protodisk. Accordingly, an angular-dependent ejecta distribution is adopted in this work, which could exhibit a clear two-segment profile.
Section 2 describes the ejecta distribution using a parameterized model. Section 3 introduces the dynamical evolution of kilonovae in the NS-powered scenario and in the radioactive decay-powered scenario. In Section 4, the main thermal and late-time radio emissions are compared under typical model parameters.
In Section 5, we apply this model to the GW\,170817 event, refitting the two-component thermal emission of AT\,2017gfo and calculating the corresponding late-time radio light curves. The conclusion and discussion are presented in Section 6.

\section{Ejecta Distribution} \label{sec:datared}

The ejecta from binary NS mergers originate from distinct mechanisms, each exhibiting significant differences in velocity, mass, electron fraction, and nucleosynthesis (see \citealt{2017LRR....20....3M}; \citealt{2021GReGr..53...59S}).
During the merger process, the tidal force ejects material predominantly within the orbital plane. The tidal dynamical ejecta reach velocities of $0.1-0.3\,c$ and have an electron fraction $Y_{\rm e}\lesssim 0.1$, reflecting the beta equilibrium of NS mater. Meanwhile, shock heating at the interface between NSs can generate polar dynamical ejecta. Due to neutrino irradiation, the polar ejecta exhibit a higher electron fraction, typically $Y_{\rm e}> 0.25$. Numerical simulations indicate that the total mass of dynamical ejecta ranges from $10^{-4}$ to $10^{-2}\,M_{\odot}$ \citep{1999A&A...341..499R, 2013ApJ...773...78B, 2013PhRvD..87b4001H}.
Another ejecta source is the outflow from the accretion disk surrounding the merger remnant. For a massive torus, the outflow mass can be comparable to or even exceed that of the dynamical ejecta, depending strongly on the lifetime of the post-merger NS \citep{2008MNRAS.390..781M, 2013MNRAS.435..502F, 2017PhRvL.119w1102S}. The disk outflows generally show a broad electron fraction distribution of $Y_{\rm e}\sim 0.1-0.4$, and the average electron fraction of the disk outflow increases with the lifetime of the post-merger NS \citep{2014MNRAS.441.3444M, 2014MNRAS.443.3134P, 2015MNRAS.450.1777K}.

Recent simulations \citep{2022MNRAS.512.1499R,2023MNRAS.520.1481Z,2025MNRAS.542..256B} indicate that the electron fraction increases with latitude, ranging from approximately 0.1 in the equatorial region to 0.4-0.5 towards the polar direction.
In this work, following \citet{2018ApJ...861L..12L}, the merger ejecta are divided into two components: the equatorial ejecta and the polar wind. The equatorial ejecta with $Y_{\rm e}<0.25$ corresponds to the red component, and the polar wind with $Y_{\rm e}>0.25$ corresponds to the blue component of the kilonova emission.  

\begin{figure}[htb]
	\includegraphics[width=\linewidth]{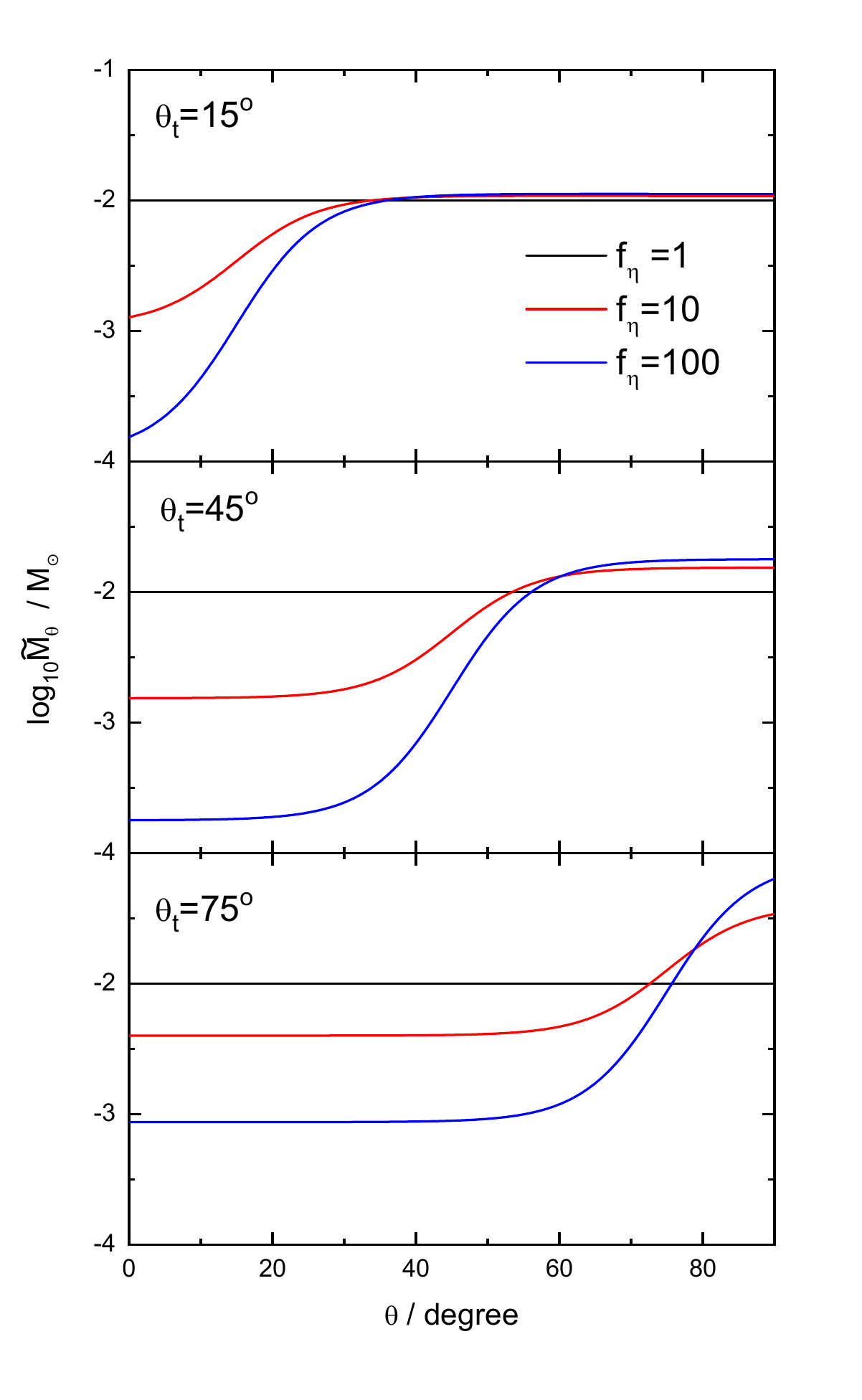}
	
	\caption{The angle-dependent equivalent mass of the ejecta. From top to bottom, the transition angles are $\theta_{\rm t} = 15^\circ$, $\theta_{\rm t} = 45^\circ$, and $\theta_{\rm t} = 75^\circ$. In each panel, the results are shown for different equatorial-to-polar mass ratios: $f_{\eta} = 1$ (black lines), $f_{\eta} = 10$ (red lines), and $f_{\eta} = 100$ (blue lines).}
	\label{fig1}
\end{figure}

The mass distribution of the ejecta is described by an angle-dependent function following \citet{2018ApJ...865L..21K} and \citet{2022Univ....8..633W},
\begin{equation}
	\eta ( \theta ) = \exp{\left[\left( 1 - \Theta \left( \theta \right) \right) \ln f_{\eta}\right]}\label{eta},
\end{equation}
where
\begin{equation}
	\Theta(\theta) = \frac{1}{1 + \exp\left[10\left(\theta - \theta_{\rm t}\right)\right]},
\end{equation}
$\theta$ is the polar angle, and $f_{\eta}$ is a parameter used to characterize the density ratio between the equatorial and polar directions. 
$\theta_{\rm t}$ is defined as the transition angle separating the high-density and low-density regions in the mass distribution. One can define an angle-dependent equivalent mass   
\begin{equation}
	\tilde{M}_{\theta}={2\eta (\theta)\over \int_{0}^{\pi} \eta (\theta)\sin\theta d\theta}M.
\end{equation}
When the polar angle $\theta>\theta_{\rm t}$, the effective mass is parameterized as $f_{\eta}$ times that in the $\theta<\theta_{\rm t}$ region, with the function employed to ensure a continuous transition between the two zones.
For the requirement of normalization condition, the effective mass satisfies $\tilde{M}_{\theta}= M_{\rm ej}$ when the parameter $f_{\eta}=1$, i.e., when the distribution of the merger ejecta is independent of the angle $\theta$. Fig.\ref{fig1} shows the equivalent mass for varying $f_{\eta}$ and $\theta_{\rm t}$. For $\theta_{\rm t} = 15^\circ$, the equatorial ejecta exhibit a wide-angle distribution and contribute most of the ejecta mass. For $\theta_{\rm t} = 45^\circ$, the ejecta masses in the equatorial and polar directions are comparable. However, the total mass of the equatorial ejecta is still larger than that of the polar component when integrated over the solid angle. For $\theta_{\rm t} = 75^\circ$, the equatorial ejecta show a disk-like geometry, and the polar component makes the dominant contribution to the total ejecta mass.

\begin{figure*}[tb]
	\includegraphics[width=2.1\columnwidth]{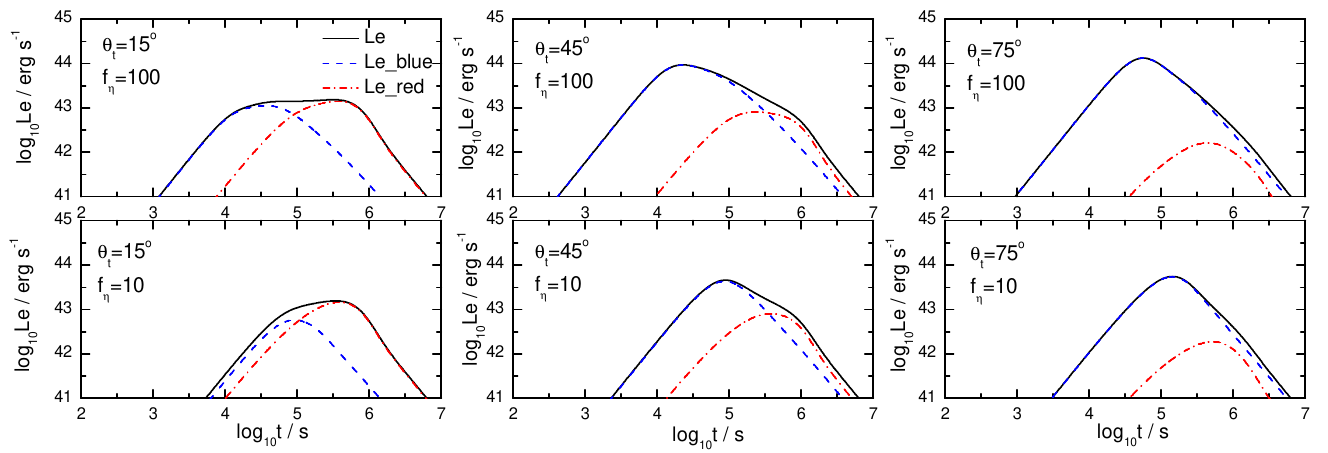}
	\caption{Light curves of the main thermal emission in the NS-powered scenario for different $\theta_{\rm t}$ and $f_{\eta}$. The blue lines and red lines correspond to the blue component and red component emissions, respectively. The central NS is assumed with $B=5\times 10^{14}\rm G$ and $P_{\rm i}=5\rm ms$. The total ejecta mass $M_{\rm{ej}} = 0.01\,M_{\odot}$, with an initial velocity $v_{\rm p,i}=0.3\,c$ for the polar wind and $v_{\rm e,i}=0.1\,c$ for the equatorial ejecta. The viewing angle is fixed at $\theta_{\rm v}=45^\circ$. The energy injection parameters are $\xi=0.3$ and $f_{\xi}=1$.}
	\label{fig2}
\end{figure*}
\begin{figure*}[tb]
	\includegraphics[width=2.1\columnwidth]{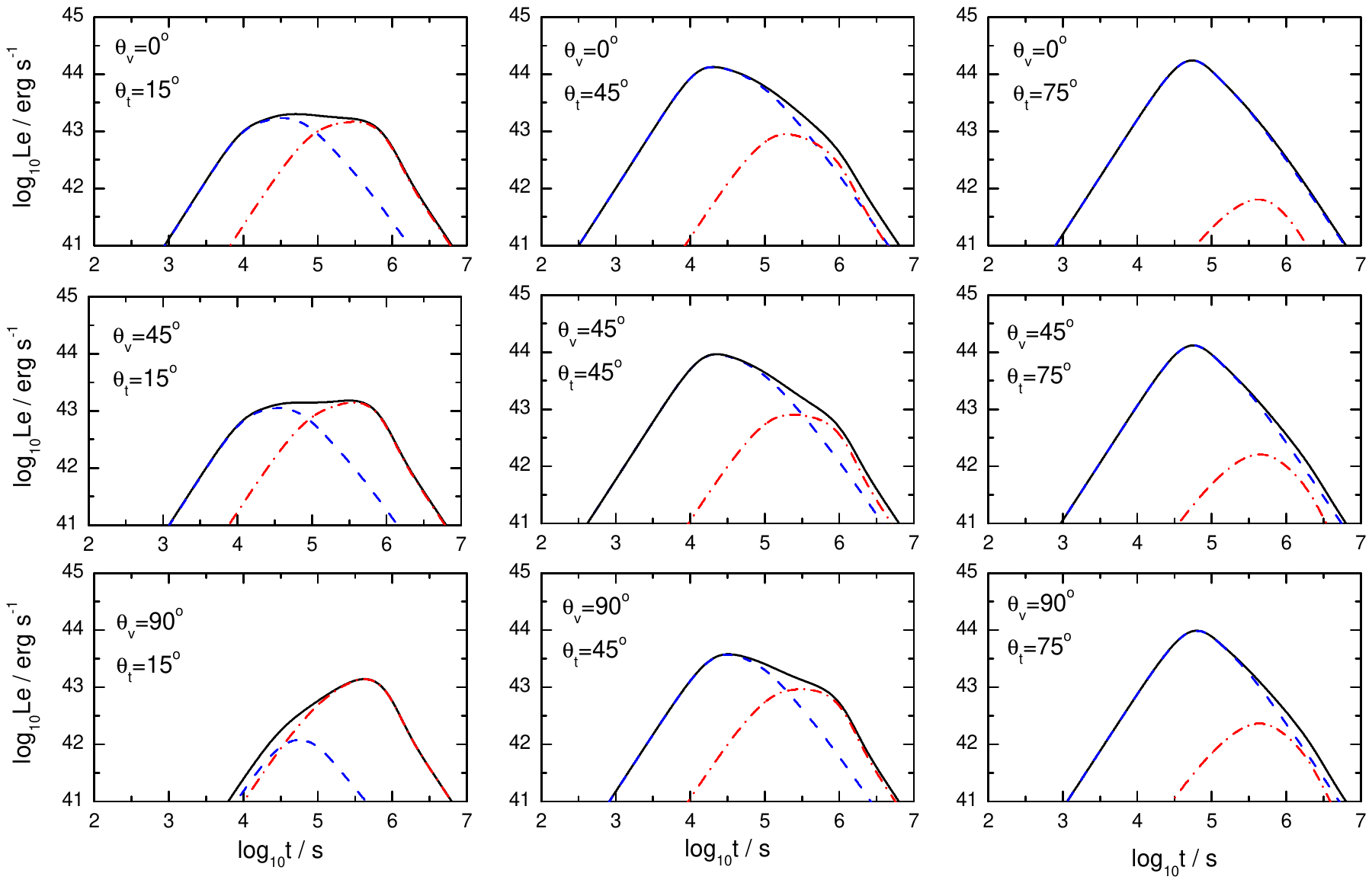}
	\caption{Light curves of the main thermal emission in the NS-powered scenario with different $\theta_{\rm t}$ and $\theta_{\rm v}$. The other parameters are the same with Fig.\ref{fig2} but with $f_{\eta}=100$.}
	\label{fig3}
\end{figure*}

\begin{figure*}[tb]
	\includegraphics[width=2.1\columnwidth]{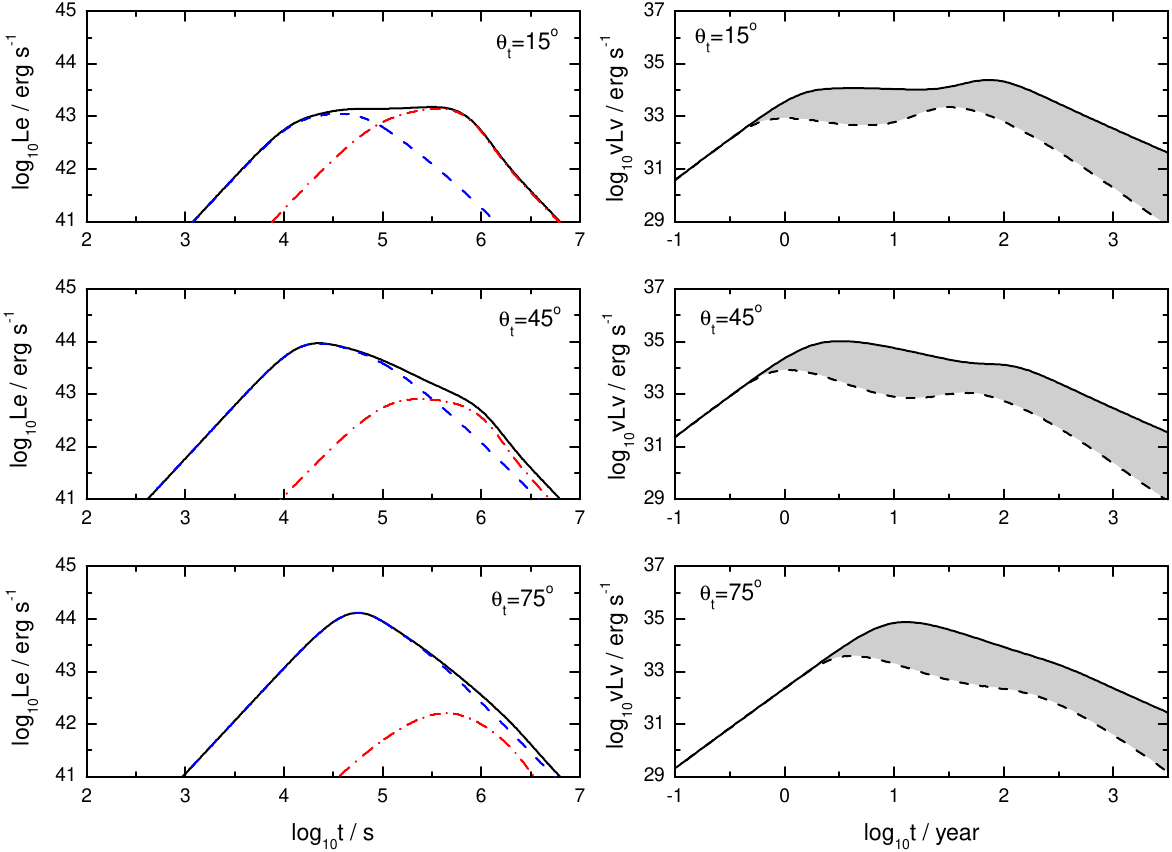}
	\caption{Light curves of the main thermal emission and the late-time radio (6GHz) emission in the NS-powered scenario. The other parameters are the same with Fig.\ref{fig3} but with a fixed $\theta_{\rm v}=45^\circ$. The parameters $n = 10^{-3}$, p = 2.2, $\epsilon_e=0.1$  and $\epsilon_B=0.1$ for the late-time radio emission. In right panels, the solid lines and dashed lines correspond to the complete-sweeping scenario and incomplete-sweeping scenario, respectively.  }
	\label{fig4}
\end{figure*}

\begin{figure*}[tb]
	\includegraphics[width=2.1\columnwidth]{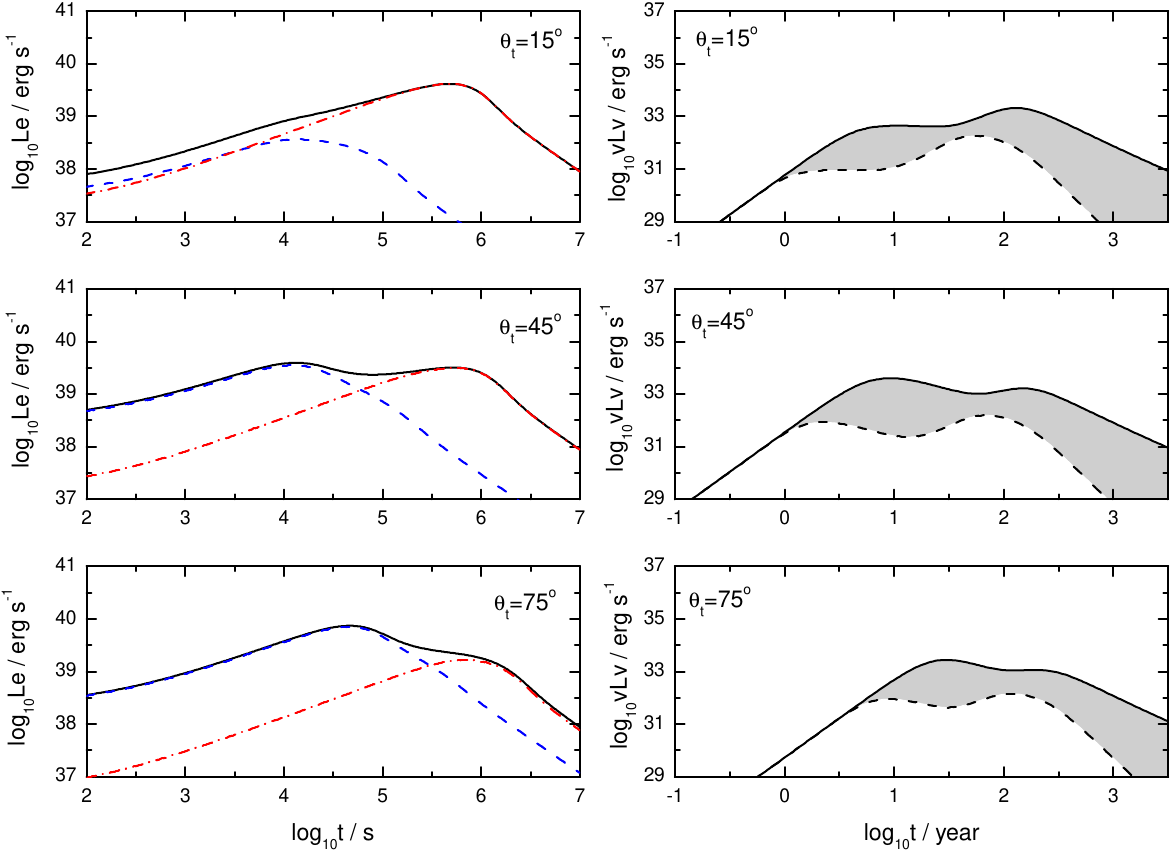}
	\caption{Light curves of the main thermal emission and the late-time radio (6\,GHz) emission in the radioactive decay scenario. The paramerters are the same as in Fig.\ref{fig4} but without the energy injection from the central NS.}
	\label{fig5}
\end{figure*}

\begin{figure*}[tb]
	\includegraphics[width=2.1\columnwidth]{ 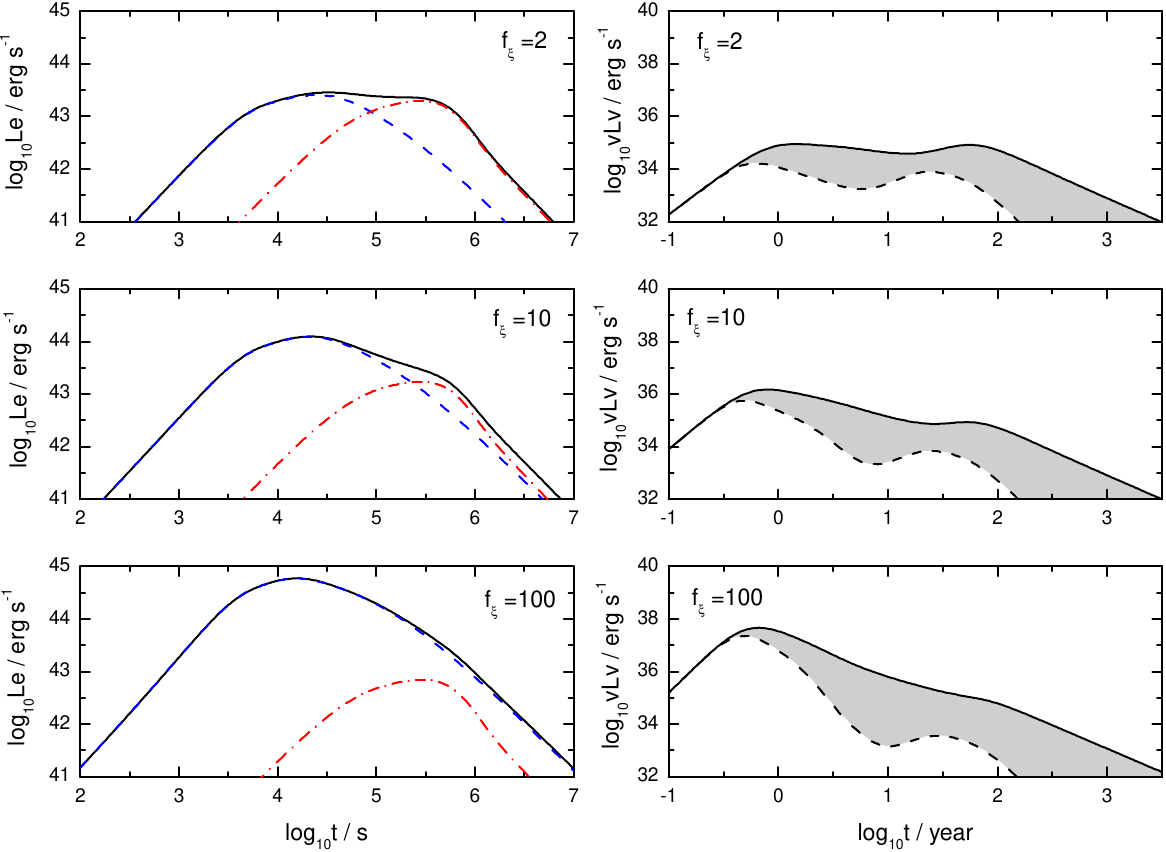}
	\caption{The same as Fig.\ref{fig4}, but with anisotropic energy injection from the central NS, for $f_{\xi}=2,\, 10,\, \& 100$. The parameters $\theta_{\rm v}=45^\circ$ and $\theta_{\rm t}=15^\circ$.  }
	\label{fig6}
\end{figure*}

\section{dynamics} 

Numerical simulation studies have demonstrated that intense magnetic field can be triggered through various magnetohydrodynamic mechanisms during and after the merger process, elevating the field strength to $\sim 10^{15}$ G-$10^{17}$ G. The primary magnetic field amplification mechanisms include:  the Kelvin-Helmholtz instability, which arises in the velocity shear layers at the merger interface \citep{2006Sci...312..719P,2008PhRvL.100s1101A,2015PhRvD..92l4034K}; the magnetorotational instability, which develops in the differentially rotating remnant star \citep{1991ApJ...376..214B,2006PhRvL..96c1101D,2018ApJ...858...52S}; and the r-mode instability driven by GW radiation \citep{2014ApJ...786L..13C, 2025MNRAS.537..650D}.
For a post-merger millisecond magnetar with an initial spin period $P_{\rm i}$, its total rotational energy is roughly
$E_{\rm{rot}} = 2 \times 10^{52} P_{\rm{i,-3}}^{-2} \, \rm{erg}$, where we adopt the cgs convention $Q_x = Q/10^x$. With a magnetic field of $B$, the spin-down luminosity of the magnetar can be described by the dipole radiation as
\begin{equation}
	L_{\rm{sd}} = L_{\rm{sd,i}} \left(1 + \frac{t}{t_{\rm{sd}}}\right)^{-2},
	\label{eq:spin_down}
\end{equation}
where $L_{\rm{sd,i}} = 10^{47} R_{\rm{s,6}}^{6} B_{14}^{2} P_{\rm{i,-3}}^{-4}\, \rm{erg}\,\rm{s}^{-1}$ is the initial spin-down luminosity and $t_{\rm{sd}} = 2 \times 10^{5} R_{\rm{s,6}}^{6} B_{14}^{2} P_{\rm{i,-3}}^{-4} \,\rm{s}$ is the spin-down timescale. For a post-merger magnetar remnant, the primary energy source of a kilonova is no longer the radioactive decay of r-process elements, but instead is the spin-down of the NS. The energy injection from a millisecond NS could be much larger than the decay power, unless the NS collapses rapidly or survives for only a short time. {In addition, recent works \citep[e.g.,][]{2024MNRAS.528.3705W,2024ApJ...976..113L} show clear anisotropic energy injection from the newborn central NS. Therefore, the equivalent spin-down luminosity can be expressed as}
\begin{equation}
	\tilde{L}_{{\rm sd},{\theta}} =
\begin{cases}
\mathcal{	C}{f_{\xi} \over f_{\xi}+1}L_{\rm sd}, & \theta<\theta_{\rm t} \\
    \mathcal{C}{1 \over f_{\xi}+1}L_{\rm sd}, & \theta>\theta_{\rm t}
\end{cases}
\end{equation}
where 
\begin{equation}
	\mathcal{C} = {4\pi (f_{\xi}+1)\over 2\pi(1-\cos \theta_{\rm t})(f_{\xi}-1)+4\pi}
\end{equation}
is the normalization factor, $f_{\xi}>1$ is the ratio of the energy injection in the polar wind to that in the equatorial ejejcta. With this expression, most of the energy is injected in the polar wind component when $f_{\xi}$ is much larger than one. This is a reasonable assumption, since the post-merger NS plays a significant role in the formation of the polar wind \citep{2013ApJ...773...78B,2013MNRAS.435..502F, 2013PhRvD..87b4001H}.

With an angle-dependent ejecta distribution, we develop the NS-powered kilonova model\citep{ 2013ApJ...776L..40Y, 2014MNRAS.441.3444M, 2018ApJ...861L..12L}. For a zero-order approximation, ignoring the lateral energy diffusion, the dynamics in each solid angle is given by 
\begin{eqnarray}
	&&\frac{d{\Gamma}_{\theta,\varphi}}{dt}=  \frac{\xi \tilde{L}_{\rm sd, \theta} + \tilde{L}_{{\rm ra},\theta} - \tilde{L}_{\rm e,{\theta,\varphi}} }{\tilde{M}_{\rm ej,\theta}c^2 + \tilde{E}'_{\rm int,{\theta,\varphi}} + 2\Gamma_{{\theta,\varphi}} \tilde{M}_{\rm sw,{\theta,\varphi}}c^2} \nonumber\\
	&&+\frac{-\Gamma_{\theta,\varphi} \mathcal{D_{\theta,\varphi}}\left(\frac{d\tilde{E}'_{\rm int,{\theta,\varphi}}}{dt'}\right) - (\Gamma_{\theta,\varphi}^2 - 1)c^2\left(\frac{d\tilde{M}_{\rm sw,{\theta,\varphi}}}{dt}\right)}{\tilde{M}_{\rm ej,\theta}c^2 + \tilde{E}'_{\rm int,{\theta,\varphi}} + 2\Gamma_{{\theta,\varphi}} \tilde{M}_{\rm sw,{\theta,\varphi}}c^2},
\end{eqnarray}
where the subscripts $\theta$ and $\varphi$ represent the direction $(\theta,\varphi)$, $\xi$ is the efficiency of the energy injection, $\tilde{E}'_{\rm int}$ is the internal energy in the comoving frame, $\tilde{M}_{\rm sw}$ and $\tilde{M}_{\rm ej}$ are the swept-up ISM mass and the ejecta mass, respectively. 
Hereafter, the tilde $\sim$ denotes the equivalent value along the direction $(\theta,\varphi)$.
{The Doppler factor is
\begin{equation}
	\mathcal{D}_{\theta,\varphi} = 1/[\Gamma_{\theta,\varphi}(1 - \beta_{\theta,\varphi}\cos\Phi )],
\end{equation} 
where $\beta_{\theta,\varphi} = \sqrt{1 - \Gamma_{\theta,\varphi}^{-2}}$ and 
$\Phi$ is the angle between the direction $(\theta,\varphi)$ and the line of sight of a distant observer. Assuming that $\theta_{\rm v}$ is the viewing angle of the distant observer, then $\Phi$ is determined by
\begin{equation}
	\cos \Phi = \sin\theta \cos\varphi \sin\theta_{\rm v} + \cos\theta \cos\theta_{\rm v},
\end{equation}
where $\varphi_{\rm v}=0$ corresponds to the observer’s azimuthal angle.}
In the comoving frame, the evolution of the internal energy can be expressed as 
\begin{equation}
	\frac{d\tilde{E}'_{\rm int,{\theta,\varphi}}}{dt'} = \xi \tilde{L}'_{\rm sd,\theta} + \tilde{L}'_{\rm ra,\theta} - \tilde{L}'_{\rm e,{\theta,\varphi}} - \mathcal{\tilde{P}'_{\theta,\varphi}} \frac{d\tilde{V}'_{\theta,\varphi}}{dt'},
\end{equation}
where the pressure $\mathcal{\tilde{P}'_{\theta,\varphi}}=\tilde{E}'_{\rm int,{\theta,\varphi}}/3\tilde{V}'_{\theta,\varphi}$. 
The volume evolves following
	\begin{equation}
		{d\tilde{V}'_{\theta,\varphi}}/{dt'}=4\pi R_{\theta,\varphi}^2 \beta_{\theta,\varphi} c,
	\end{equation}
	together with 
	\begin{equation}
		{dR_{\theta,\varphi}}/{dt}={\beta_{\theta,\varphi} c \over 1- \beta_{\theta,\varphi}\cos\Phi },
	\end{equation}
which {includes the equal-time surface effect}.
Note that the energy injection from the radioactive decay is defined in the co-moving frame as 
\begin{eqnarray}
	\tilde{L}'_{\rm ra,\theta} &=& \frac{\tilde{L}_{\rm{ra,\theta}}}{\mathcal{D_{\theta,\varphi}}^2} \nonumber\\
	&=& 4 \times 10^{49} \tilde{M}_{\rm{ej},\theta,-2} \left[ \frac{1}{2} - \frac{1}{\pi} \arctan \left( \frac{t' - t'_0}{t'_\sigma} \right) \right]^{1.3} \, {\rm erg\, s^{-1}}
\end{eqnarray}
where $t'_0\sim 1.3\rm\, s$ and $t'_\sigma\sim 0.11\rm\, s$ \citep{2012MNRAS.426.1940K}.
The equivalent diffusion luminosity in the comoving frame is determined by
\begin{equation}
	\tilde{L}'_{{\rm e},{\theta,\varphi}} =
	\begin{cases}
		{{E}'_{{\rm int},{\theta,\varphi}}} c \over \tau_{\theta,\varphi} R_{\theta,\varphi}/\Gamma_{\theta,\varphi}, & \tau>1 \\
		{{E}'_{{\rm int},{\theta,\varphi}}} c \over R_{\theta,\varphi}/\Gamma_{\theta,\varphi}, & \tau<1
	\end{cases}
\end{equation}
where $\tau_{\theta,\varphi} =  \kappa_{\theta} (\tilde{M}_{\rm{ej,\theta}}/{V'_{\theta,\varphi}} ) ( R_{\theta,\varphi}/{\Gamma_{\theta,\varphi}})$ is the optical depth in the $\theta$ direction, and $\kappa_\theta$ is the opacity. 
The opacity follows an angle-dependent distribution similar to that of the ejecta mass, $\kappa_\theta= \eta ( \theta ) \kappa_{0}$, but with a fixed $f_{\eta,\kappa}=5$ and  $\kappa_{0}=1$. Then, the ejecta is lanthanide-free with $\kappa\simeq 1$ in the polar direction and lanthanide-rich with $\kappa\simeq 5$ in the equatorial direction \citep{2013ApJ...774...25K,2018ApJ...861L..12L,2018ApJ...852..109T}. {The same treatment is also applied to the initial velocity of the ejecta, where $v_{\theta}= \eta ( \theta ) v_{\rm i}$, $f_{\eta, v}=v_{\rm e,i}/v_{\rm p,i}$ and $v_{\rm i} = v_{\rm p, i}$, with $v_{\rm p, i}$ and $v_{\rm e, i}$ being the initial velocities in the polar and equatorial directions, respectively}. In the observer's frame, the angle-dependent luminosity reads $\tilde{L}_{{\rm e},{\theta,\varphi}}=\mathcal{{D}}^2_{\theta,\varphi} \tilde{L}_{{\rm e},{\theta,\varphi}}^{\prime}$. {Then, one can obtain the effective
temperature for the distant observer, 
\begin{eqnarray}
 \tilde{T}_{\rm eff,{\theta,\varphi}}= \left(\tilde{L}_{{\rm e},{\theta,\varphi}} \over  4\pi R_{\rm ph,{\theta,\varphi}}^2 \sigma \right)^{1/4},
\end{eqnarray}
where $\sigma$ is the Stefan-Boltzmann constant and $R_{\rm ph}$ is the photospheric radius. The condition $\tau>1$ is usually satisfied at the peak timescale of the kilonova light curves, so the photospheric radius has $R_{\rm ph}\sim R$ except at very late times.}
{According to Lambert's cosine law, the contribution from the direction $(\theta,\varphi)$ to the observed luminosity relates to $\cos\Phi $. Then the total main thermal emission of the kilonova can be obtained by integrating over the solid angle, which is
\begin{eqnarray}
	L_{\rm e} =\int\limits_{\cos \Phi >0} \tilde{L}_{{\rm e},{\theta,\varphi}}{\cos\Phi  ds_{\theta,\varphi} \over \pi R_{\theta,\varphi}^2} \label{Le}
\end{eqnarray}
where $ds_{\theta,\varphi}=R_{\theta}^2 \sin \theta d\theta d\varphi$.}

Under the energy injection from the post-merger NS, the ejecta are accelerated to higher velocities compared to the radioactive decay scenario. 
Such fast-moving ejecta drive a stronger external shock interacting with the ISM, emitting much brighter late-time radio signals via synchrotron 
radiation \citep{2016ApJ...831..141F, 2016ApJ...819L..22H}. If the medium is completely swept up by the expanding ejecta, the mass of the swept-up medium increases with radius as 
\begin{equation}
	\frac{d{M}_{\rm{sw,{\theta,\varphi}}}}{dR_{\theta,\varphi}} = 4\pi R_{\theta,\varphi}^2 n m_{\mathrm{p}}.\label{msw}
\end{equation}
where $n$ is the number density of the surrounding medium. However, if the ISM is predominantly composed of neutral rather than ionized hydrogen, the merger ejecta can hardly sweep up the medium at late times \citep{2024ApJ...961..201L}. In this case, a suppression factor of 
$(1-e^{-{R_{{\theta,\varphi}}/l}})$ should be included on the right-hand side of Eq.\ref{msw}, where $l$ is the mean free path of particles traversing the ejecta.
In such an incomplete-sweeping scenario, the mass of swept-up medium decreases as the ejecta expand.

The energized electrons in the shocked medium lose their energy via synchrotron emission. The equivalent angle-dependent radio spectral luminosity can be expressed as \citep{2020ApJ...890..102L, 2024ApJ...961..201L}
\begin{equation}
	{\tilde{L}}_{\nu,{\theta,\varphi}}={(1+z)N_{{\rm e},{\theta,\varphi}}}\frac{m_{\rm e}c^2 \sigma_{\rm T}\Gamma_{\theta,\varphi}\beta_{{\rm m},{\theta,\varphi}}^2 B_{\theta,\varphi}'}{3e} \left( \frac{\nu_{\rm obs}}{\nu_{{\rm m},{\theta,\varphi}}}\right)^{-(p-1)/2} .
\end{equation}
where $N_{{\rm e},{\theta,\varphi}}=M_{{\rm sw},{\theta,\varphi}}/m_{\rm p}$ is the number of shocked electrons. 
The characteristic synchrotron frequency of the electrons is given by \rm{$\nu_{{\rm m},{\theta,\varphi}}={3 \Gamma_{\theta,\varphi} \gamma_{\rm m,{\theta,\varphi}}^2 \beta_{\rm m,{\theta,\varphi}}^2 e B_{\theta,\varphi}'}/{4 \pi m_e c}$}, and $\gamma_{\rm m,{\theta,\varphi}} = \left( {(p - 2) m_p}/{(p - 1) m_e} \right) \epsilon_e (\Gamma_{\rm {\theta,\varphi}} - 1) + 1$ is the minimum Lorentz factor of the shocked electrons. The parameters $p=2.2$ and $\epsilon_e=0.1$ are adopted here. Note that the observed frequency always have $\nu_{\rm obs}>\nu_{\rm m},\nu_{\rm a}$ in most realistic scenarios (Nakar \& Piran 2011). The magnetic field strength in the comoving frame of the external shock is  
$B'_{\theta,\varphi}= \sqrt{8\pi {\epsilon}_{\rm B}(4\Gamma_{\theta,\varphi}+3)(\Gamma_{\theta,\varphi}-1){{n}}_{\rm }m_{\rm p}c^2}$,
where ${\epsilon}_{\rm B}$ is the energy fraction deposited in the magnetic field. {Similarly, the integrated radio spectral luminosity is} 
\begin{equation}
	L_{\nu}=\int\limits_{\cos \Phi >0} {\tilde{L}}_{\nu,{\theta,\varphi}}{\cos\Phi  ds_{\theta,\varphi} \over \pi R_{\theta,\varphi}^2}\label{Lv}.
\end{equation}
From Eq.\ref{Le} and Eq.\ref{Lv}, it can be seen that both the main thermal emission and the late-time radio emission are connected to the specific distribution of the ejecta. Therefore, they are expected to exhibit some potential correlation in their light curves.
Additionally, it should be noted that the beaming effect is not taken into account here since the merger ejecta can hardly be accelerated to ultra-relativistic velocity. 

\section{emission} 

{In most merger scenarios, off-axis viewing is more common from an observational perspective.  Therefore, we first fix the viewing angle at $45^{\circ}$ and calculate the kilonova emission. Fig.\ref{fig2} shows the resulting kilonova light curves in the NS-powered scenario for different values of $\theta_{\rm t}$
and $f_{\eta}$. For a two-segment anisotropic ejecta distribution, the main thermal emission observed exhibits a double-peaked structure. The results indicate that the double-peaked structure is not very pronounced for $f_{\eta}=10$; however, when $f_{\eta}=100$, a clear double-peaked profile emerges. In other words, for a distinct double-peaked structure to emerge in the light curve, the density of the polar wind must be approximately two orders of magnitude lower than that of the equatorial ejecta.}

{For more general cases, Fig.\ref{fig3} shows that the contributions of the polar and equatorial ejecta to the observed luminosity depend on the different viewing angle $\theta_{\rm v}$ and the transition angle $\theta_{\rm t}$ . Under the same parameters, when the viewing angle is closer to the polar direction and $\theta_{\rm t}$ is much larger than $45^{\circ}$, the polar wind component contributes more, leading to a more luminous early blue-component. Conversely, when the viewing angle is closer to the equatorial plane and $\theta_{\rm t}$ is much smaller than $45^{\circ}$, the equatorial ejecta contribution dominates, resulting in a more prominent late red-component emission. In particular, when $\theta_{\rm v}=90^{\circ}$ and $\theta_{\rm t}=15^{\circ}$, only a small fraction of the radiation from the polar wind can be observed, which is consistent with Lambert’s cosine law. 
Except for this extreme case, the light curves exhibit only a weak dependence on the viewing angle and show a similar evolution. This is physically reasonable, as a distant observer always receives radiation from nearly half of the ejecta's surface. Although the largest contribution comes from the ejecta surface directed toward the observer, with a symmetric ejecta distribution, most of the integrated results for different $\theta_{\rm v}$  do not show a significant difference in the light curves.
Due to the cosine factor, only the ejecta moving nearly perpendicular to the line of sight contribute negligibly. Consequently, the luminosity of the early blue component is significantly reduced for $\theta_{\rm v}=0^{\circ}$ and $\theta_{\rm t}=75^{\circ}$. }

{With an NS as the central engine, both the main thermal and the late-time radio emission are powered by the central NS. These two types of emission are plotted together in Fig.\ref{fig4}. In this plot, the viewing angle is fixed at $45^{\circ}$ and different values of $\theta_{\rm t}$ are adopted to characterize the anisotropy of the ejecta. As mentioned above, a clear double-peaked profile emerges not only in the main thermal emission, but also in the late-time radio light curves. In particular, it shows a high degree of consistency in the temporal evolution of the  light curves of these two types of emission. This connection fundamentally originates from the fact that both types of emission depend sensitively on the ejecta distribution.
For comparison, we also show the radioactive decay‑powered scenario in Fig.\ref{fig5}. Although the light curves of the radioactive decay‑powered scenario also exhibit a double-peaked profile, the consistency between the main thermal and late-time radio light curves is weaker than that in the  NS‑powered scenario. The energy injection from the central NS is independent of the ejecta distribution,  whereas the energy injection from radioactive decay is coupled with the ejecta distribution. For a total ejecta mass of $0.01\, M_{\odot}$ , the peak luminosity from the radioactive decay-powered scenario is usually lower than that of the NS-powered scenario. For the polar wind component, due to its low mass fraction, the energy supplied by radioactive decay is insufficient to accelerate the ejecta to higher velocities. In this case, the double-peaked profile in the light curves more directly reflects the difference in the initial velocity of the ejecta. In contrast, the energy injection from the central NS can effectively accelerate the polar wind ejecta, producing significant differences in the dynamics of the polar wind and the equatorial ejecta.}

\begin{figure}[t]
	\includegraphics[width=\linewidth]{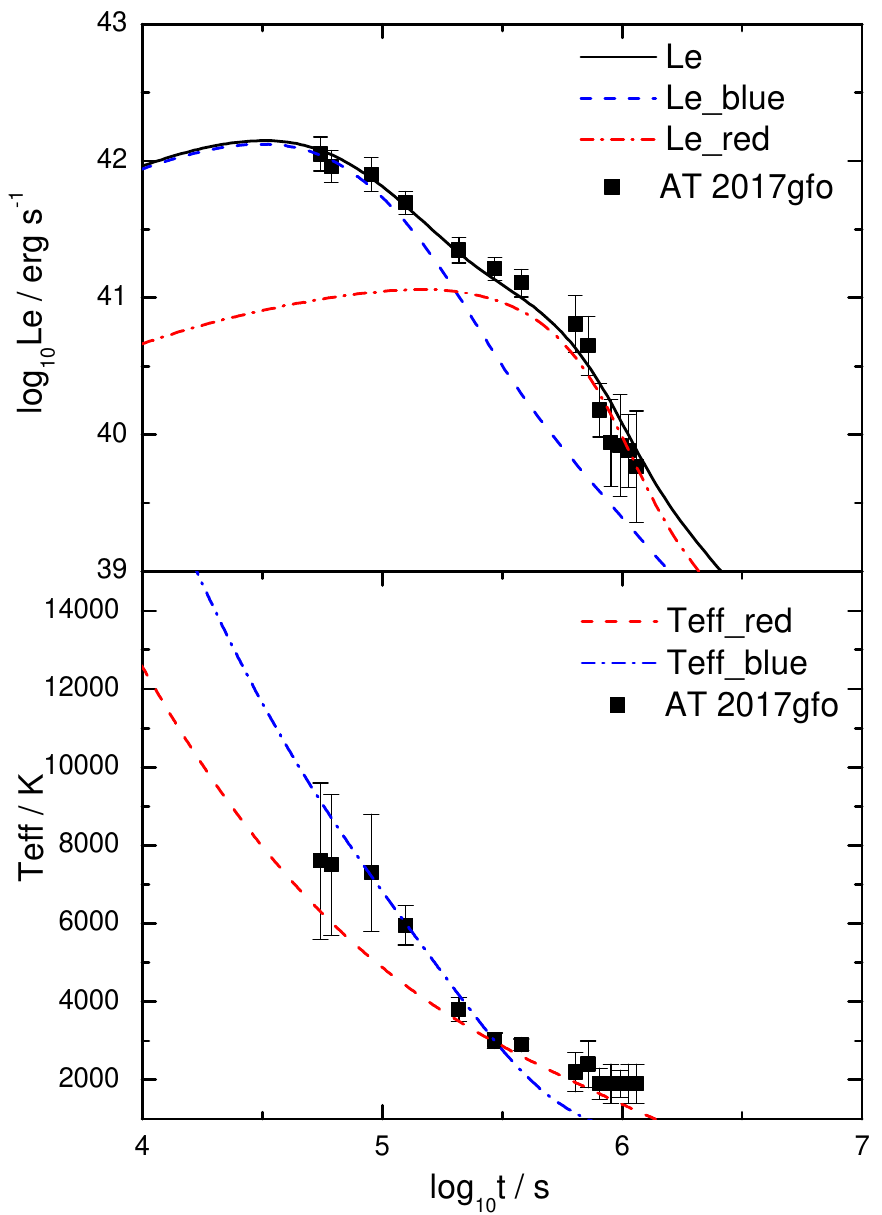}
	
	\caption{Light curves and effective temperatures of AT\,2017gfo. The effective temperatures correspond to the polar direction ($\theta_{\rm t}=0^\circ$, blue) and the equatorial direction ($\theta_{\rm t}=90^\circ$, red), respectively. The fitting parameters are: $\theta_{\rm v} = 30^\circ$, $\theta_{\rm t} = 15^\circ$, $M_{\rm{ej}} = 0.0085\, M_{\odot}$, $f_{\eta}=5$, $v_{\rm p,0}=0.32\,c$, $v_{\rm e,0}=0.12\,c$, $B=4.2\times 10^{15}\rm G$, $P_{\rm i}=6.6\,\rm ms$, $\xi = 0.3$, and $f_{\xi}=10$. The observation data are from \citet{2017Natur.551...75S}.}
	\label{fig7}
\end{figure}

\begin{figure}[t]
	\includegraphics[width=\linewidth]{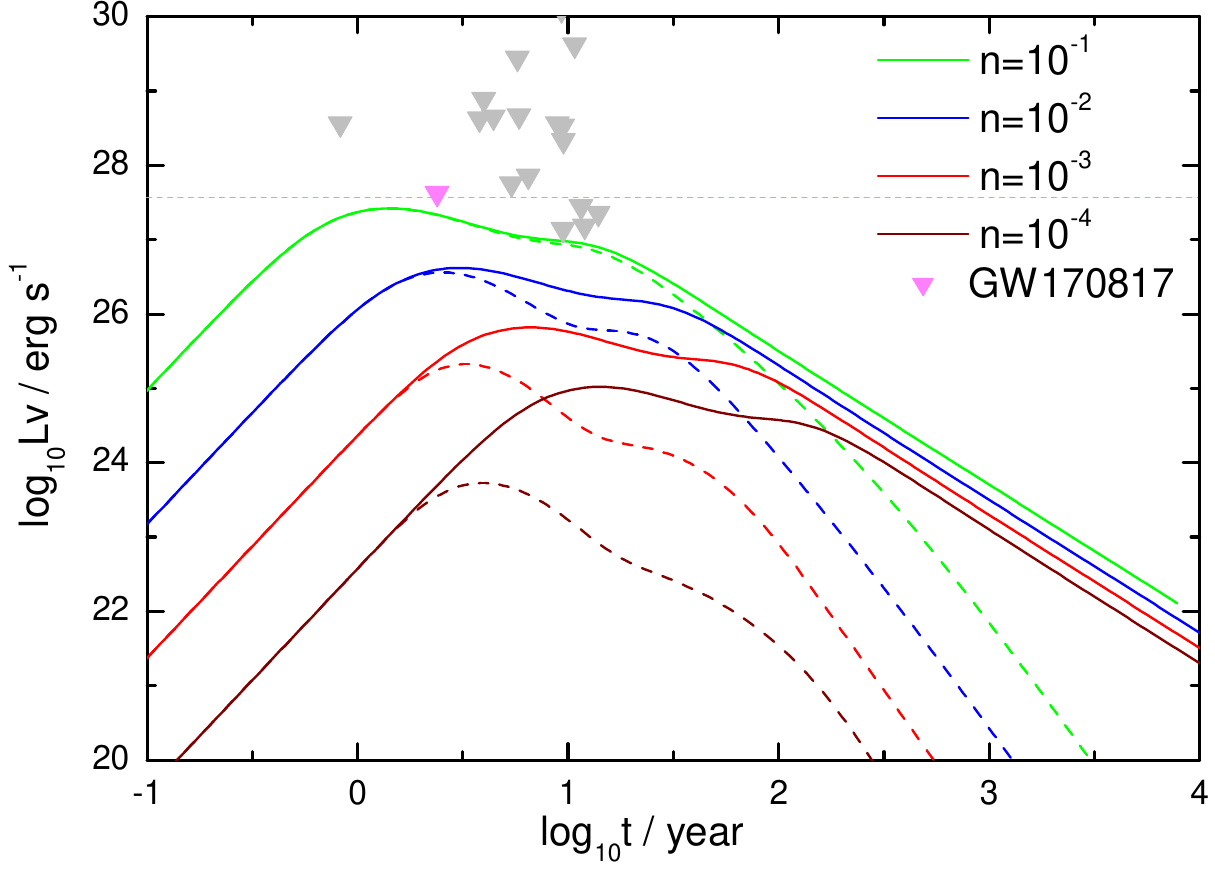}
	
	\caption{Potential late-time radio light curves of the GW\,170817 event for different ISM densities. The solid and dashed lines correspond to the complete-sweeping and incomplete-sweeping scenarios, respectively. The model parameters are derived from Fig.~\ref{fig7}. The upper limit for GW\,170817 is marked in pink, while other limits for short GRB samples are shown in black for comparison. The observation data are taken from \citet{2021MNRAS.500.1708R}.  }
	\label{fig8}
\end{figure}

{The energy released from an NS is usually assumed to be isotropic. However, recent simulation work indicates that more NS energy is injected into the polar direction \citep{2024MNRAS.528.3705W}. Accordingly, Fig.\ref{fig6} presents the light curves for different values of $f_{\xi}$ with a fixed $\theta_{\rm t}=15^\circ$. It can be seen that the stronger the anisotropy of the energy injection (corresponding to a larger $f_{\xi}$), the more energy is injected into the polar wind. Consequently, the blue component emission from the polar wind arrives earlier with a higher luminosity. In case that most of the energy is injected into the polar direction, the luminosity of the equatorial ejecta is significantly reduced, and in extreme cases, it is completely covered by the emission from the polar wind. Although the energy injection from the NS enhances the overall anisotropic signature, the main thermal and late-time radio emission still exhibit a consistent evolution in their light curves, which is determined by the underlying dynamical mechanisms governing both components.	}

In addition, the late-time radio light curves for the incomplete-sweeping scenario are plotted as dashed lines in Fig.\ref{fig4}, Fig.\ref{fig5} and Fig.\ref{fig6}, which correspond to the case of neutral hydrogen medium (see \citealt{2024ApJ...961..201L}). More generally, the ISM is a mixture of ionized and neutral hydrogen, with an ionization fraction $\in(0,1)$. Consequently, the more plausible late-time radio light curve should lie within the shaded regions as shown in these figures.

Whether powered by an NS or by radioactive decay, the light curves of the main thermal emission and the late-time radio emission exhibit a certain degree of consistency. This is essentially caused by the anisotropic distribution of the merger ejecta.
On one hand, we can infer the peak flux and the light-curve shape of the late-time radio emission from the observed early thermal emission; on the other hand, 
the combination of these two types of emission can be used to constrain the properties of the merger ejecta as well as the central engine, reducing the model degeneracy.

\section{GW\,170817}

GW\,170817 is the first binary NS merger event detected by gravitational-wave observatories. The optical transient AT\,2017gfo, associated with GW\,170817, provides important observational constraints on the merger remnant, the mass and velocity of the ejecta, and the nucleosynthesis process. 
Based on the fitting of the multi-band afterglow of GRB\,170817A, the viewing angle is inferred to be $\lesssim 30^{\circ}$\citep{2021ApJ...909..114N}. We therefore use a fixed viewing angle $\theta_{\rm v}=30^{\circ}$ in the following fitting
to reduce the degrees of freedom.
Fig.~\ref{fig7} presents the light curve of the main thermal emission of AT\,2017gfo, fitted with the anisotropic ejecta distribution model in the NS-powered scenario. 
Since the anisotropic energy injection from the central NS may play a significant role in the formation of the polar wind ejecta, 
, $\theta_{\rm t}=15^\circ$ is adopted in our fitting to characterize a narrow distribution of the polar wind ejecta. As Fig.\ref{fig7} shows,
a good fit is achieved with an ejecta mass of $0.0085\, M_{\odot}$. This is consistent with estimates from other works \citep{2018ApJ...860...57A, 2018ApJ...861L..12L, 2018ApJ...861..114Y}, which suggested that the ejecta mass could be significantly lower than $0.06 M_{\odot}$ if AT\,2017gfo were powered  by an NS or magnetar. 
The anisotropy parameters $f_{\eta }=5$, and $f_{\xi }=10$ indicate that both the ejecta distribution and the energy injection should be anisotropic.
The density of the polar wind is much lower than that of the equatorial dynamical ejecta, and more energy is injected into the polar wind ejecta. 
Consequently, the polar wind component contributes most of the earlier blue emission of AT\,2017gfo, while the equatorial ejecta account for the later red component.

Furthermore, based on the parameters derived from fitting the main thermal emission light curve, we calculate the corresponding late-time radio light curve, as shown in Fig.\ref{fig8}. A comparison with the upper limits reveals that, for an ISM density $n < 10^{-1} \rm{cm}^{-3}$, a persistent late-time radio signal is not expected to be detectable with current observational sensitivities. Our fitting results show that a central NS with an initial period of $6.6\,\rm ms$ is required to explain the thermal emission of AT\,2017gfo . It is significantly longer than the Keplerian period of a newborn NS, implying that most of the rotational energy of a newly born NS has been carried away by other efficient braking mechanisms.  
With such a central engine, the energy injected into the ejecta is only on the order of $10^{51} \rm erg$, which is insufficient to produce a strong late-time radio signal. Therefore, the non-detection of the late-time radio signals from GW\,170817 is, in a sense, consistent with theoretical expectations.

\section{Conclusions and discussion}

This study investigates the impact of the anisotropic mass distribution of the ejecta from binary NS mergers on the main thermal and late-time radio emissions of kilonovae. 
The angular momentum and the mass ratio of the binary system can lead to significant anisotropy in the angular distribution of the merger ejecta.
We use a parameterized distribution model to characterize the configuration of the anisotropic ejecta.
With a central NS or radioactive decay as the engine, we numerically solve the dynamical equations and obtain the light curves of the main thermal emission and the potential late-time radio emission.
The results demonstrate that the ejecta distribution significantly affects the light-curve profile. 
In addition, the anisotropy in the energy injection from the central NS is discussed. Using such an anisotropic model, we calculate the late-time radio emission from GW\,170817 based on the fitting parameters from AT\,2017gfo and compare the resulting radio light curves with the observational constraints.
The findings can be summarized as follows:
\begin{enumerate}[(i)]
	\item Regardless of whether the engine is a central NS or radioactive decay, the light curves of the main thermal emission and the late-time radio emission exhibit a certain degree of consistency in their temporal evolution. This connection arises primarily from the fact that the anisotropic distribution of the ejecta plays a significant role both in energy diffusion and the interaction with the ISM.
	\item  Based on this consistency, one can use the light curves of the main thermal emission to constrain the flux and shape of the late-time radio light curves.
	\item  If AT\,2017gfo is indeed powered by a central NS, the associated late-time radio signal from GW\,170817 would likely fall below the detection threshold of current observations for typical parameters.
	\item The connection between the main thermal and late-time radio emission provides a dual-consistency check for the magneta model, offering further evidence in support of an NS remnant.
\end{enumerate}

Finally, the most critical issue for the magnetar model is determining where the enormous rotational energy of the newborn NS ultimately goes. Neither the multi-band afterglows of short GRBs, the isotropic kilonova thermal emission, nor the X-ray plateaus of long GRBs have revealed radiation at the $10^{52} \rm erg$ level corresponding to a massive NS with Keplerian rotation. A plausible channel for this substantial energy loss is gravitational-wave radiation \citep{2016PhRvD..93d4065G,2022MNRAS.513.1365X, 2023MNRAS.522.4294Y,  2025ApJ...986...14H}. A newly formed NS can develop quadrupole moments due to various mechanisms \citep[e.g.,][]{1995ApJ...442..259L,  1996A&A...312..675B, 1998ApJ...502..708A, 2015ApJ...798...25D}, leading to significant secular gravitational-wave emission. Given the unknown waveform and the sensitivity limits of current gravitational-wave detectors, existing observations cannot reliably constrain how much energy is carried away by post-merger gravitational-wave radiation. Regardless, future detector upgrades and more observational data will significantly advance our understanding of the post-merger remnant \citep{2017ApJ...851L..16A,2020LRR....23....3A, 2024ChPhB..33h0401C, 2024PhRvD.110h3016G, 2025ApJ...986...14H, 2025arXiv251121941K}.

\begin{acknowledgements}
The authors sincerely thank the reviewer for the insightful and constructive report. This work is supported by the National Natural
Science Foundation of China (grant No. 12303044).
\end{acknowledgements}


\bibliographystyle{aasjournalv7}




\end{document}